\documentclass{iaus}
\usepackage{graphics}
  \checkfont{eurm10}
  \iffontfound
    \IfFileExists{upmath.sty}
      {\typeout{^^JFound AMS Euler Roman fonts on the system,
                   using the 'upmath' package.^^J}%
       \usepackage{upmath}}
      {\typeout{^^JFound AMS Euler Roman fonts on the system, but you
                   dont seem to have the}%
       \typeout{'upmath' package installed. iaus.cls can take advantage
                 of these fonts,^^Jif you use 'upmath' package.^^J}%
      }
  \else
  \fi

  \checkfont{msam10}
  \iffontfound
    \IfFileExists{amssymb.sty}
      {\typeout{^^JFound AMS Symbol fonts on the system, using the
                'amssymb' package.^^J}%
       \usepackage{amssymb}%

      }{}
  \fi

  \IfFileExists{amsbsy.sty}
    {\typeout{^^JFound the 'amsbsy' package on the system, using it.^^J}%
     \usepackage{amsbsy}}
    {\providecommand\boldsymbol[1]{\mbox{\boldmath $##1$}}}

\newcommand\etal{\mbox{\textit{et al.}}}

\title[The Interplay among Black Holes, Stars and ISM in Galactic 
       Nuclei]{     Multiwavelength View of SDSS Galaxies       }

\author[M. Obri\'{c} {\it et al.\/}]%
{M. Obri\'{c}$^{1,2}$
\v{Z}. Ivezi\'c$^{1,3}$, R.H. Lupton$^{1}$, G. Kauffmann$^{4}$, G.R. Knapp$^{1}$, J.E. Gunn$^{1}$,
D. Schlegel$^{1}$, M.A. Strauss$^{1}$, S. Anderson$^{3}$, et al.}

\affiliation{$^1$Princeton University, Princeton, USA, 
$^2$Kapteyn Institute, Groningen, The Netherlands,
$^3$University of Washington, Seattle, USA,
$^4$Max-Planck-Inst. f\"ur Astroph., Garching, Germany}

\pubyear{2004}
\volume{222}
\pagerange{1--8}
\date{?? and in revised form ??}
\jname{The Interplay among Black Holes, Stars and ISM \\in Galactic Nuclei}
\editors{Th. Storchi Bergmann, L.C. Ho \& H.R. Schmitt, eds.}
\begin{document}

\maketitle

\begin{abstract}
We summarize the detection rates at wavelengths other than optical for $\sim$99,000 
galaxies from the Sloan Digital Sky Survey (SDSS) Data Release 1 ``main''
spectroscopic sample. The analysis is based on positional cross-correlation
with source catalogs from ROSAT, 2MASS, IRAS, GB6, FIRST, NVSS and WENSS 
surveys. We find that the rest-frame UV-IR broad-band galaxy SEDs form a remarkably 
uniform, nearly one parameter, family. As an example, the SDSS $u$ and $r$ 
band data, supplemented with redshift, can be used to predict $K$ band magnitudes 
measured by 2MASS with an rms scatter of only 0.2 mag; when measurement uncertainties 
are taken into account, the astrophysical scatter appears not larger than $\sim$0.1 mag.
\end{abstract}

{\bf \large \hskip -0.14 in 1. The fractions of SDSS galaxies detected at other wavelengths}

We report initial results from a program aimed at the characterization of 
multiwavelength properties of galaxies detected by SDSS. The details about the
positional cross-correlation method, and analysis of matched samples, will be 
described elsewhere (Obri\'{c} et al. 2004); here we summarize the detection
rates and demonstrate a tight correlation between the UV and IR colors
of galaxies. 

Table 1 lists the detection rates for galaxies from SDSS ``main'' spectroscopic 
sample (a flux-limited sample, $r_{Pet}<17.77$, for details see Strauss et al.
2002) by ROSAT, 2MASS, IRAS, GB6, FIRST, NVSS and WENSS surveys. Galaxies 
are separated into those with and without emission lines (using line measurements
and classification criteria from Kauffmann et al. 2003), and emission line
galaxies are further separated, using Baldwin-Phillips-Terlevich diagram, 
into ``star-forming'', ``AGNs'', and ``unclassified''. The detection rate of 
SDSS galaxies is below 10\% in all surveys except 2MASS.

\begin{table}
\begin{center}
\vskip -0.2in
\caption{{\bf The detection rates of SDSS galaxies at other wavelengths}}
\begin{tabular}{cccccccc}
Catalog & Wavelength  & Total & No emission & Emission & AGN
& SF & Unclassified\\
%\hline  
% SDSS (N    )& U-V & 99088   & 50900 & 43281 & 15264 & 9742 & 18275\\
$\Sigma_{SDSS}$ (\#/deg$^2$) & UV-IR & 72.86 & 39.38 & 33.48 & 11.81 & 7.54 & 13.44 \\
2MASS XSC [\%]  & near-IR & 38.06& 35.72 & 39.64 & 63.79 & 10.69 & 34.00\\
IRAS  [\%]  & far-IR & 2.25 &  0.29  &  3.73 &  3.73 &  2.31 & 4.50 \\
GB6   [\%]  & 6 cm & 0.37 &  0.44 &  0.26 &  0.43 &  0.17 & 0.16\\
FIRST [\%]  & 20 cm & 3.86 &  2.76 &  4.68 &  8.06 &  0.98 & 3.40\\
NVSS  [\%]  & 20 cm & 7.78 &  7.06 &  7.85 &  9.54 &  5.68 & 7.50\\
WENSS [\%]  & 92 cm   & 0.61 & 0.77 &  0.40 &  0.61 &  0.25 & 0.30\\
ROSAT [\%]  & X-ray   & 0.27 &  0.27 &  0.20 &  0.37 &  0.10 & 0.098\\
 \end{tabular}
  \label{tab}
 \end{center}
\end{table}

\begin{figure}
\centering
%\resizebox{3.7cm}{!}{\includegraphics{obric_figure.ps} } 
\resizebox{2.7cm}{!}{\includegraphics{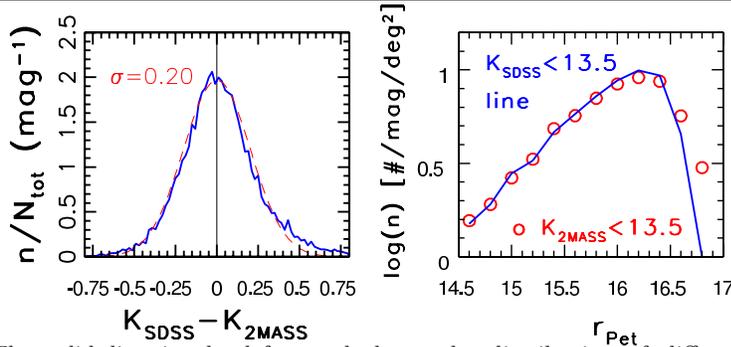} } 
\caption{The solid line in the left panel shows the
distribution of differences between SDSS-predicted
and 2MASS-measured $K$ band flux. The dashed line 
is a Gaussian with $\sigma=0.20$ mag. The right 
panel shows the $r$ band counts of SDSS-predicted
(line) and 2MASS-measured (symbols) galaxies with 
$K<13.5$ (2MASS XSC completeness limit).}
\label{fig}
\end{figure}

{\bf \large \vskip 0.1in \hskip -0.14 in 2. Predicting 2MASS K flux from UV/optical SDSS flux}

Galaxies form a nearly one-dimensional sequence in various optical
color-color diagrams (the correlation among colors is especially tight
for rest-frame colors, with a scatter of only $\sim$0.03 mag perpendicular 
to the locus, Smol\v{c}i\'{c} et al. 2004). We report here that this one-dimensionality 
of galaxy spectral energy distributions (SEDs) extends to 2MASS wavelengths: 
{\it it is possible to predict 2MASS $K$ band flux with an rms of 0.2 mag 
using SDSS $u$ and $r$ band measurements\footnote{We use UV/visual fluxes 
($u$ and $r$ bands) to predict near-IR flux because this is the ``hardest'' 
wavelength combination with most astrophysical implications (according to 
``common wisdom", such a relationship should not be very accurate due to 
the effects of starbursts and dust extinction); predicting, for example,
2MASS $J$ band flux from SDSS $z$ band flux is trivial because these two 
bands are adjacent in wavelength space.}.} 

The SDSS-based $K$ magnitude prediction is determined from $K_{SDSS} =
r_{Pet} - (r-K)$, where $r_{Pet}$ is the SDSS $r$ band Petrosian
magnitude (Strauss et al. 2002), and $(r-K)$ is a best fit to the 
observed $r-K$ colors for SDSS/2MASS galaxies.
%(without correcting for the difference beween AB and Vega magnitudes). 
It depends on $u-r$ color
(essentially a position along the Hubble sequence), and on redshift, 
%to account for K-correction effects\footnote{We use default 2MASS magnitudes
%(Jarrett et al. 2000), and SDSS model colors.}:\\ 
to account for 
K-correction effects (we use default 2MASS magnitudes and SDSS model colors):
\newpage
\phantom{only2pages...}
\vskip -0.3in
\hskip -0.14in 
%K-correction effects (we use default 2MASS magnitudes and SDSS model colors):\\ 
%\begin{equation}
%(r-K)_{fit} = A+B\,(u-r)+C\,(u-r)^2+D\,(u-r)^3+E\,z+F\,z^2
%\end{equation}
\vskip -0.12in
$\phantom{mala} (r-K) = A+B\,(u-r)+C\,(u-r)^2+D\,(u-r)^3+E\,z+F\,z^2  \phantom{mirelajezlocesta} (2.1) \\$
\vskip -0.12in
\hskip -0.14in
where (A, B, C, D, E, F)=(1.115, 0.940, -0.165, 0.00851, 4.92, -9.10),
and $z$ is redshift. This fit predicts 2MASS $K$ band magnitudes with an 
rms scatter of 0.20 mag, which depends on neither color nor redshift, and 
is nearly Gaussian (see Figure 1).

To correct for aperture and resolution effects 
between SDSS and 2MASS, that presumably depend on galaxy profile, or nearly 
equivalently on galaxy color (Strateva et al. 2000), we add to $(r-K)$
($0.496-0.154\,R_{50}^z$) for galaxies with $u-r<2.22$ and
($0.107-0.045\,R_{50}^z$) for redder galaxies, where $R_{50}^z$ is the
radius containing 50\% of Petrosian flux in the $z$ band.  
This correction has a negligible effect on rms scatter, and only removes a correlation
of $K_{SDSS}-K_{2MASS}$ residuals with galaxy size. 

The median residuals, as a function of $u-r$ and $z$, do not exceed 0.03 mag.
 The rms scatter decreases to 0.15 mag at the bright end
($K<12$). Given typical measurement errors in $u$, $r$, $R_{50}^z$ and
$K$, we conservatively conclude that the true astrophysical scatter of
$K$ band magnitudes predicted from the blue part of SED is less than $\sim$0.1
mag. Similarly, the relation $(J-K)=2.172\,z + 0.966$, where $z$ is
redshift, predicts $J-K$ measured by 2MASS with an rms scatter of 0.11
mag (0.07 mag at the bright end), and no significant residuals with
respect to $K$, $u-r$ and redshift. These tight correlations demonstrate
remarkable one-dimensionality of galaxy spectral energy distributions from 
UV to IR wavelengths.

\end{document}